%% file: main_ICC_workshop.tex
\documentclass[conference]{IEEEtran}
\IEEEoverridecommandlockouts

\usepackage{cite}
\usepackage{amsmath,amssymb,amsfonts}
\usepackage{algorithmic}
\usepackage{graphicx}
\usepackage{textcomp}
\usepackage{xcolor}
\usepackage{algorithm}
\usepackage{stmaryrd}
\usepackage{color} 
\usepackage{graphicx,psfrag, cite}
\usepackage[acronym]{glossaries}
\newcommand{\newac}{\newacronym}
\newcommand{\ac}{\gls}
\newcommand{\Ac}{\Gls}
\newcommand{\acpl}{\glspl}

\makeglossaries
\input{JCgroup_acronym.tex}
\usepackage{amsthm}
\providecommand{\theoremname}{Theorem}
\providecommand{\remarkname}{Remark}
\providecommand{\definitionname}{Definition}
\providecommand{\lemmaname}{Lemma}
\providecommand{\propositionname}{Proposition}
\theoremstyle{plain}
\newtheorem{thm}{\protect\theoremname}
\theoremstyle{remark}

\theoremstyle{definition}

\theoremstyle{plain}
\newtheorem{lem}[thm]{\protect\lemmaname}
\theoremstyle{plain}
\newtheorem{prop}[thm]{\protect\propositionname}
\theoremstyle{plain}

\def\BibTeX{{\rm B\kern-.05em{\sc i\kern-.025em b}\kern-.08em
    T\kern-.1667em\lower.7ex\hbox{E}\kern-.125emX}}

\begin{document}
\newac{psd2}{PSD}{power spectral density}
\newac{slf}{SLF}{spatial loss function}
\newac{btd}{BTD}{block term tensor decomposition}
\newac{tps}{TPS}{thin plate spline}
\newac{nmse}{NMSE}{normalized mean squared error}
\newac{cp}{CP}{cyclic prefix}
\newac{cfo}{CFO}{carrier frequency offset}
\newac{sic}{SIC}{successive interference cancellation}
\newac{ml}{ML}{maximum likelihood}
\newac{crlb}{CRLB}{Cramer-Rao lower bound}
\newac{nr}{NR}{new radio}
\newac{ss}{SS}{synchronization signal}
\newac{fim}{FIM}{Fisher information matrix}
\newac{ofdm}{OFDM}{orthogonal frequency division multiplexing}
\newac{ris}{RIS}{reconfigurable intelligent surfaces}
\newac{ssb}{SSB}{SS block}
\newac{glrt}{GLRT}{generalized likelihood ratio test}
\newac{ue}{UE}{user equipment}
\newac{lls}{LLS}{linear least squares}
\newac{usrp}{USRP}{universal software radio peripheral}
\newac{nmae}{NMAE}{normalized mean absolute error}
\makeatother

\title{Joint Channel and CFO Estimation From Beam-Swept Synchronization Signal Under Strong Inter-Cell Interference}

\author{\IEEEauthorblockN{Bowen~Li\IEEEauthorrefmark{1}\IEEEauthorrefmark{2}, Junting~Chen\IEEEauthorrefmark{2}\IEEEauthorrefmark{3}, Nikolaos~Pappas\IEEEauthorrefmark{1}}
\IEEEauthorblockA{\IEEEauthorrefmark{1} School of Science and Engineering and Shenzhen Future Network of Intelligence Institute, \\ The Chinese University of Hong Kong (Shenzhen), China}
\IEEEauthorblockA{\IEEEauthorrefmark{2} Department of Computer and Information Science, Link{\"o}ping University, 58183, Link{\"o}ping, Sweden}
\IEEEauthorblockA{\IEEEauthorrefmark{3} Corresponding author}

\thanks{This work was supported in part by National Key R\&D Program of China under Grant 2024YFB2907500, Guangdong Basic and Applied Basic Research Foundation 2024A1515011206, and Shenzhen Science and Technology Program No. KJZD20230923115104009, in part by the European Union (6G-LEADER) under 101192080, and ELLIIT.}
}
\maketitle

\begin{abstract}
Complete awareness of the wireless environment, crucial for future intelligent networks, requires sensing all transmitted signals, not just the strongest. A fundamental barrier is estimating the target signal when it is buried under strong co-channel interference from other transmitters, a failure of which renders the signal unusable. This work proposes a \ac{ml}-based cross-preamble estimation framework that exploits \ac{cfo} constancy across beam-swept \ac{ss}, coherently aggregating information across multiple observations to reinforce the desired signal against overwhelming interference. \ac{crlb} analysis and simulation demonstrate reliable estimation even when the signal is over a thousand times weaker than the interference. A low-altitude radio-map case study further verifies the framework's practical effectiveness.
\end{abstract}


\section{Introduction}

\label{sec:intro}

The paradigm of wireless networks is evolving towards complex, multi-point architectures where awareness of the entire radio environment, not just the strongest signal, is critical \cite{ZarCav:J08,GesHanHuaSha:J10,NgoAshYanLar:J17,MaeOzdGuvSic:M23,ZenCheXuWu:J24,SunChe:J24}. For instance, cell-free systems require channel information from numerous access points to enable cooperation \cite{ZarCav:J08,GesHanHuaSha:J10,NgoAshYanLar:J17}. Likewise,
high-resolution radio map construction needs to characterize all signals to provide a complete environmental picture \cite{MaeOzdGuvSic:M23,ZenCheXuWu:J24,SunChe:J24}. A unifying theme across these systems is the necessity of processing weak signals in the presence of strong ones. This creates a challenge: weak signals of interest are often buried under severe co-channel interference, and the critical first step of \ac{cfo} estimation fails, rendering them invisible to the network.

To estimate \ac{cfo}, blind methods exploit inherent signal redundancy, such as the \ac{cp}, for estimation without dedicated overhead \cite{LinPho:J16,SinKumMajSat:J25}. However, their reliance on short data segments provides limited averaging gain, rendering them unreliable in low \ac{sinr} and strong interference conditions. Training-based methods offer improved robustness by utilizing known preambles or pilots as a clean reference \cite{TsaHuaCheYan:J13,SalNasMehXia:J17,TunRivGarMel:J23}.
However, practical synchronization signals are never perfectly orthogonal \cite{StoBes:J03,HuaWanYanZou:J14}, leading to cross-correlation leakage that allows strong interfering signals to corrupt the estimation of weaker ones. Even dedicated iterative techniques like \ac{sic} \cite{MorKuoPun:J07}, designed specifically for this challenge, often fall short, as residual estimation errors from strong signals are typically still powerful enough to completely mask the weak signals.

We tackle this challenge by exploiting a structural property of modern training bursts. Within a short coherence block, multiple preambles exhibit heterogeneous per-sequence (beam-dependent) gains, while the \ac{cfo} is common across the burst. For example, in 5G \ac{nr} \ac{ssb} structure \cite{3gpp:ts38211,TunRivGarMel:J23,YonSawNagSuy:J25}: each \ac{ssb} has a distinct, beam-dependent power gain, yet all \acpl{ssb} share a common \ac{cfo} over a short duration ($\approx 5$ ms); in cell-free networks \cite{ZarCav:J08,GesHanHuaSha:J10,NgoAshYanLar:J17}: the same pilot from a user is received by multiple \acpl{bs} through different channels, so the received gains differ across sites, whereas the user-induced \ac{cfo} is identical at all receivers. Our approach leverages this structure, proposing a cross-preamble estimation across the entire burst to reinforce the weak signal against strong co-channel interference.

This paper develops a principled framework for weak-signal synchronization. It establishes the performance limits, showing that it is not the instantaneous \ac{sinr} that matters, but the aggregate \ac{sinr} across all observations. Our key contributions are:
\begin{itemize}
\item We develop an \ac{ml}-based cross-preamble \ac{cfo} estimator that exploits a shared \ac{cfo} across multiple preambles and, via \ac{crlb} analysis, show that the estimation error is inversely proportional to the aggregate \ac{sinr} over all preambles and proportional to the \ac{cfo} magnitude. Therefore, an iterative refinement algorithm, pre-compensation, and \ac{sic}, followed by re-estimation, is proposed to estimate the channel and \ac{cfo}.
\item Simulations demonstrate reliable \ac{cfo} estimation down to $-30\,\mathrm{dB}$ \ac{sinr} and an approximate $6$ dB improvement in channel estimation at low \ac{sinr}. A low-altitude radio-map field test further confirms the framework's practical effectiveness.
\end{itemize}

\section{System Model}

\label{sec:sys_model}

\begin{figure}
\begin{centering}
\includegraphics[width=1\columnwidth]{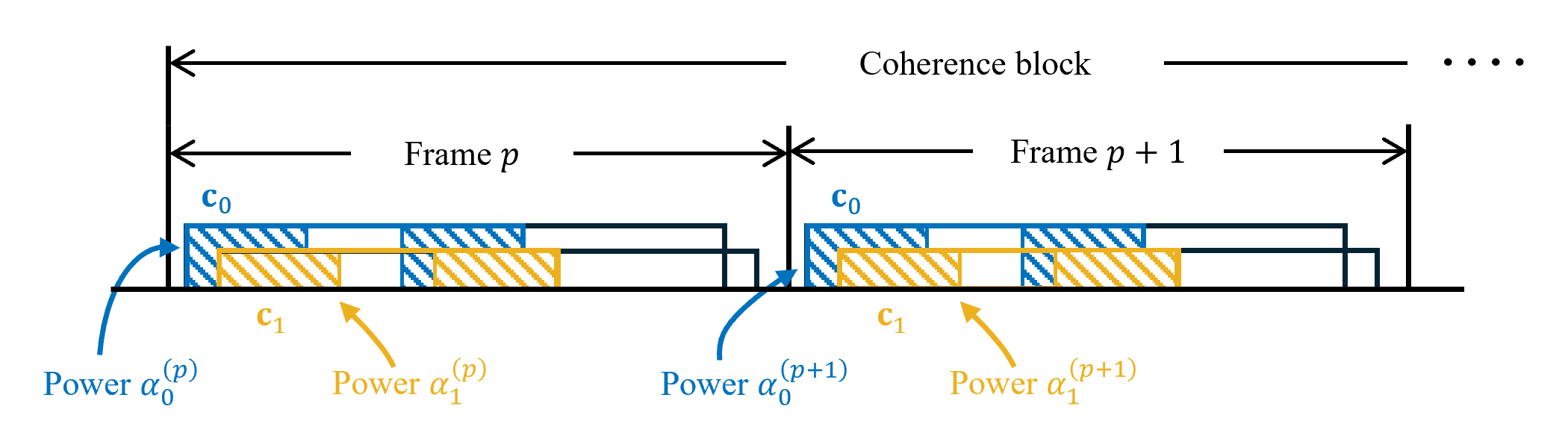}
\par\end{centering}
\caption{\label{fig:coherence_block_system}Preamble Structure.}
\end{figure}

Consider a multi-transmitter system where a single receiver is within the coverage of $K$ \acpl{bs}, indexed by $\mathcal{K}\in\{0,1,\cdots,K-1\}$. Each \ac{bs} periodically broadcasts a signal burst containing $P$ distinct preamble frames, indexed by $\mathcal{P}\in\{0,1,\cdots,P-1\}$. All preamble frames contain two identical training sequences, as shown in Fig.~\ref{fig:coherence_block_system}. In addition, all preambles within one coherence block share the same timing and frequency offsets, while each preamble frame experiences a distinct channel.

\subsection{Preamble Structure}

Let $\mathcal{C}$ denote the set of all length-$N$ training sequences. The preamble transmitted by the $k$th \ac{bs} can be expressed as
\begin{equation}
\mathbf{c}_{k}=\mathbf{c}_{k,0}\oplus\mathbf{0}\oplus\mu_{k}\mathbf{c}_{k,1}\label{eq:def_frame}
\end{equation}
where $\mathbf{c}_{k,0},\mathbf{c}_{k,1}\in\mathcal{C}$ are the selected training sequences; individual sequences may be reused across different \ac{bs}, but the ordered pair ($\mathbf{c}_{k,0},\mathbf{c}_{k,1}$), and thus the preamble $\mathbf{c}_{k}$ is unique to \ac{bs} $k$; $\mu_{k}\in\mathbb{R}^{+}$ is a scaling factor that represents the power difference between the two training sequences; and $\mathbf{0}$ is a zero-padding vector of fixed length $\tau_{0}$, corresponding to the timing offset between the two training signals.

The cross-correlation between two training sequences $\mathbf{c}_{k},\mathbf{c}_{l}\in\mathcal{C}$
with timing offset $\tau$ and frequency offset $\omega$ is given by
\begin{equation}
r_{k,l}\left(\tau,\omega\right)\triangleq\frac{1}{N}\sum_{m=0}^{N-1}c_{k}\left[m\right]c_{l}^{*}\left[m-\tau\right]e^{-j\omega m}.\label{eq:def_ac}
\end{equation}
Because training sequences are not perfectly orthogonal, we model the cross-correlation for any mismatch ($k\neq l$) or nonzero delay ($\tau\neq0$) as a zero-mean circular complex Gaussian with a variance that decays with the sequence length $N$. For the matched case ($k=l$ and $\tau=0$), the statistic reduces to the sequence\textquoteright s autocorrelation $r_{k}(\omega)$. As a result, the cross-correlation can be expressed as 
\begin{equation}
r_{k,l}\left(\tau,\omega\right)=\begin{cases}
r_{k}\left(\omega\right) & k=l,\tau=0\\
\mathcal{CN}\left(0,\sigma_{\text{c}}^{2}/N\right) & \text{otherwise}
\end{cases}\label{eq:dis_ac}
\end{equation}
where $\sigma_{\text{c}}^{2}$ captures sequence-family sidelobes
and implementation impairments (e.g., filtering, residual multipath),
and the magnitude of autocorrelation $|r_{k}(\omega)|$ follows a
Dirichlet profile; $|r_{k}(0)|=1$ and although oscillatory, its overall
trend decreases as $\left|\omega\right|$ increases, \emph{e.g.},
for Zadoff--Chu sequences, $|r_{k}(\omega)|=|\sin(N\omega/2)/(N\sin(\omega/2))|$
\cite{HuaWanYanZou:J14}.

\subsection{Transmission Model}

For the $k$th \ac{bs}, denote $\tau_{k}$ and $\omega_{k}$ as the
group timing and frequency offsets respectively, and denote $\alpha_{k}^{(p)}$ denote the complex channel gain in the $i$th frame. Then, the received signal at time $m$ can be expressed as\footnote{The proposed algorithm can be naturally extended to frequency-selective fading by estimating the multipath gains and delays. However, this work focuses on the single-path (frequency-flat) case, which is sufficient for the low-altitude scenarios considered herein.}
\begin{equation}
y\left[m\right]=\sum_{k=0}^{K-1}\sum_{p=0}^{P-1}\alpha_{k}^{\left(p\right)}c_{k}\left[m-pN_{\text{f}}-\tau_{k}\right]e^{-j\omega_{k}m}+\nu\left[m\right]\label{eq:def_rx_signal}
\end{equation}
for all $m\in\{0,1,\cdots,N_{\text{o}}-1\}$, where $N_{\text{f}}$
is the frame length and $\nu[m]\sim\mathcal{CN}(0,\sigma_{\text{n}}^{2})$
denotes additive white Gaussian noise. The parameter $N_{\text{o}}$
represents the total observation length under consideration. Without
loss of generality, we assume that the received signal length is sufficiently
large to accommodate all reference signals, and that training sequences
from different frames do not overlap. Moreover, $c_{k}[m]=0$ for
$m\notin\{0,\cdots,N-1\}$.

For notational convenience, we collect the received samples into a vector 
$
\mathbf{y}\triangleq\left[y\left[0\right],y\left[1\right],\cdots,y\left[N_{\text{o}}-1\right]\right]^{\text{T}}.
$

In this work, the initial tasks of active \ac{bs} detection and delay estimation are assumed to be complete, consistent with established \ac{glrt} methods \cite{Kay:B98v2}. The scope is therefore narrowed to estimating the remaining parameters for the active set $\mathcal{K}$:  \acpl{cfo} $\boldsymbol{\omega}\triangleq[\omega_k]_{k\in\mathcal{K}}$, scaling factors $\boldsymbol{\mu}\triangleq[\mu_k]_{k\in\mathcal{K}}$, and per-frame channel gains $\boldsymbol{\alpha}\triangleq[\alpha_k^{(p)}]_{k\in\mathcal{K},\,p\in\mathcal{P}}$.
\[
\underset{\boldsymbol{\omega},\boldsymbol{\alpha},\boldsymbol{\tau},\boldsymbol{\mu}}{\text{maximize}}\quad\mathbb{P}\left(\mathbf{y}|\boldsymbol{\omega},\boldsymbol{\alpha},\boldsymbol{\mu}\right).
\]

This joint estimation problem is challenging due to strong parameter coupling, the inherent nonlinearity of \ac{cfo} estimation, and imperfect orthogonality of the training sequences. 
\section{Cross-Preamble Estimation}
In this section, we first reformulate channel estimation as a linear problem and solve it via the classical \ac{lls} estimator \cite{KAy:B93}. Second, we perform a single-preamble analysis and derive a separate \ac{cfo} estimator. Then, we extend the analysis across preambles, leveraging the burstwise shared \ac{cfo} to obtain a cross-preamble \ac{cfo} estimator.
\subsection{Channel Estimator}
For each frame of the received signal, denoted as $\mathbf{y}^{(p)}\triangleq y\left[m+pN_{\text{f}}\right]_{m\in\{0,1,\cdots,N_{\text{f}}-1\}}\in\mathbb{C}^{N_{\text{f}}\times1}$,
where $p\in\mathcal{P}$, then the $p$th received signal can be written
as
\[
\mathbf{y}^{(p)}= \underbrace{e^{j\omega (p-1)N_{\text{f}}} \boldsymbol{\Omega}\left(\boldsymbol{\omega}\right) \varodot \mathbf{C}\left(\boldsymbol{\tau}\right)}_{\boldsymbol{A}^{(p)}}\boldsymbol{\alpha}^{\left(p\right)}+\mathbf{v}^{(p)}
\]
where $\mathbf{C}\left(\boldsymbol{\tau}\right)=[\mathbf{0}_{\tau\times1};\mathbf{c}_{k};\mathbf{0}]_{k\in\mathcal{K}}\in\mathbb{C}^{N_{\text{f}}\times K}$,
$\boldsymbol{\Omega}\left(\boldsymbol{\omega}\right)=[\Omega_k(\omega_k)]_{k\in\mathcal{K}}\in\mathbb{C}^{N_{\text{f}}\times K}$, $\Omega_k(\omega_k)=[1;e^{j\omega_{k}};\cdots;e^{j\omega_{k}(N_{\text{f}}-1)}]$, 
$\boldsymbol{\alpha}^{(p)}=[\alpha_{k}^{(p)}]_{k\in\mathcal{K}}\in\mathbb{C}^{K\times1}$,
and $\mathbf{v}^{(p)}=[v[m]]_{m\in\{0,\cdots,N_{\text{f}}-1\}}\in\mathbb{C}^{N_{\text{f}}\times1}$.

As a result, the received signals can be expressed as 
\[
\mathbf{y} = \underbrace{\text{blkdiag}\left(\boldsymbol{A}^{(0)},\cdots,\boldsymbol{A}^{(P-1)}\right)}_{\boldsymbol{A}(\boldsymbol{\tau},\boldsymbol{\omega})} \boldsymbol{\alpha} + \boldsymbol{v}. \label{eq:y_matrix_form_1}
\]
Then, according to the \ac{lls} solution, the channel
gain $\boldsymbol{\alpha}$ can be estimated
by 
\begin{equation}
\hat{\boldsymbol{\alpha}}=\boldsymbol{A}(\boldsymbol{\tau},\boldsymbol{\omega})^{\dagger}\mathbf{y}.\label{eq:est_alpha}
\end{equation}

In addition, the received signals $\mathbf{y}$ can be expressed as 
\[
\boldsymbol{A}_0 \boldsymbol{\alpha} + 
\underbrace{\left[\boldsymbol{A}^{(0)} \text{diag}\left( \boldsymbol{\alpha}^{\left(0\right)}\right);\cdots; \boldsymbol{A}^{(P-1)} \text{diag}\left( \boldsymbol{\alpha}^{\left(P-1\right)}\right) \right]}_{\boldsymbol{B}}\boldsymbol{\mu} +\boldsymbol{v}.
\]
where $\boldsymbol{A}_i = \text{blkdiag}(\boldsymbol{A}_i^{(0)},\cdots,\boldsymbol{A}_i^{(P-1)})$, $\boldsymbol{A}_i^{(p)}=e^{j\omega (p-1) N_{\text{f}}}$ $\boldsymbol{\Omega}(\boldsymbol{\omega}) \varodot \mathbf{C}_i\left(\boldsymbol{\tau}\right)$, $\mathbf{C}_i\left(\boldsymbol{\tau}\right) = [\mathbf{0}_{(\tau+\tau_\text{c}i)\times1};\mathbf{c}_{k,i};\mathbf{0}]_{k\in\mathcal{K}}$, and $\tau_{\text{c}}=N+\tau_{0}$.
Then, according to the \ac{lls} solution, the scaling factor $\boldsymbol{\mu}$ can be estimated
by 
\begin{equation}
\hat{\boldsymbol{\mu}}=\boldsymbol{B}^{\dagger}\left(\mathbf{y}-\boldsymbol{A}_0\right).\label{eq:est_mu}
\end{equation}

\subsection{Separate Analysis and Separate \ac{cfo} Estimator}
For an arbitrary frame $p$ of the received signal, we define the cross-correlation between the received signal $\mathbf{y}$ and the two training sequences $\mathbf{c}_{k,0}$ and $\mathbf{c}_{k,1}$ as
\begin{equation}
r_{y,k}^{p,i}\left[\tau\right]\triangleq\frac{1}{N}\sum_{m=\tau}^{\tau +N-1}y\left[m+pN_{\text{f}}+i\tau_{\text{c}}\right]c_{k,i}^{\text{H}}\left[m-\tau\right]\label{eq:def_x_r_yk}
\end{equation}
where $k\in\mathcal{K}$, $p\in\mathcal{P}$, and $i\in\{0,1\}$.

Based on the cross-relation of the training sequences, the distribution of $r_{y,k}^{p,i}\left[\tau\right]$ is summarized as
\begin{lem}\label{lem:x_r_yk}
For any $k\in\mathcal{K}$, $p\in\mathcal{P}$,
and $i\in\{0,1\}$, if $\tau=\tau_k$, the cross-correlation $r_{y,k}^{p,i}\left[\tau_k\right]$
follows 
\begin{equation}
r_{y,k}^{p,i}\left[\tau_k\right]\sim\alpha_{k}^{(p)}r_{k}\left(\omega_{k}\right)\text{\ensuremath{\left(\mu_{k}e^{-j\omega_{k}\tau_{\text{c}}}\right)}}^{i}+\mathcal{CN}\left(0,\sigma_{k,p,i}^{2}\right)\label{eq:rst_x_r_yk}
\end{equation}
where the $\sigma_{k,p,i}^{2}$ is the averaged interference plus
noise 
\[
\sigma_{k,p,i}^{2}=\frac{\sum_{q\neq k,q\in\mathcal{K}}\left(\mu_{k}^{i}\alpha_{q}^{\left(p\right)}\sigma_{c}\right)^{2}+\sigma_{n}^{2}}{N}.
\]
\end{lem}

\begin{proof}
(Sketch) Substituting the received signal expression (\ref{eq:def_rx_signal}) into the cross-correlation definition (\ref{eq:def_x_r_yk}), the resulting summation contains three parts: 

i) Desired Signal: The auto-correlation yields the deterministic mean $\alpha_{k}^{(p)}r_{k}(\omega_{k})(\mu_{k}e^{-j\omega_{k}\tau_{\text{c}}})^{i}$.

ii) Interference: Cross-correlations with other transmitters $q\neq k$ produce a zero-mean complex Gaussian interference term with variance $\sum_{q\neq k,q\in\mathcal{K}}(\mu_{k}^{i}\alpha_{q}^{(p)}\sigma_{c})^{2}/N$.

iii) Noise: Correlation with noise contributes an additional zero-mean complex Gaussian noise term with variance $\sigma_{n}^{2}/N$.

Combining these contributions yields (\ref{eq:rst_x_r_yk}).
\end{proof}

Lemma~\ref{lem:x_r_yk} indicates that the ratio between the two
correlation outputs, $r_{y,k}^{p,0}$ and $r_{y,k}^{p,1}$,
reveals the exponential term associated with the \ac{cfo}, which leads to the following separate \ac{cfo} estimator: 
\begin{equation}
\hat{\omega}_{k}^{\text{sep}}=-\frac{1}{\tau_{\text{c}}}\angle\left(\mathbb{E}\left[r_{y,k}\left[p,1\right]\right]/\mu_{k} / \mathbb{E}\left[r_{y,k}\left[p,0\right]\right]\right).\label{eq:single_est_omega}
\end{equation}
However, its variance depends on the per-sequence \ac{sinr} $\delta^2_{k,p,i}$ and is easily overwhelmed by strong co-channel signals, which motivates the cross-preamble estimator below.

\subsection{Cross-Preamble \ac{cfo} Estimator}

Let the set of correlation variables associated with \ac{cfo} $\omega_{k}$
as $\mathbf{r}_{y,k}=[r_{y,k}^{p,i}\left[\tau_k\right]]_{p\in\mathcal{P},i\in\{0,1\}}$,
and define the cross-preamble likelihood as $\mathbb{P}(\mathbf{r}_{y,k}|\boldsymbol{\omega})$.
Since preambles from different frames are non-overlapping, the correlation
outputs are conditionally independent given $\omega_{k}$, yielding
\[
\mathbb{P}\left(\mathbf{r}_{y,k}|\boldsymbol{\omega}\right)=\prod_{p\in\mathcal{P}}\prod_{i\in\{0,1\}}\mathbb{P}\left(r_{y,k}^{p,i}\left[\tau_k\right]|\omega_{k}\right).
\]

According to Lemma~\ref{lem:x_r_yk}, each correlation output $r_{y,k}^{p,i}\left[\tau_k\right]$
is Gaussian distributed, and thus the log-likelihood reduces to 
\begin{equation}
\sum_{p\in\mathcal{P}}\sum_{i\in\left\{ 0,1\right\} }-\frac{\left|r_{y,k}^{p,i}\left[\tau_k\right]-\alpha_{k}^{(p)}r_{k}\left(\omega_{k}\right)\text{\ensuremath{\left(\mu_{k}e^{-j\omega_{k}\tau_{\text{c}}}\right)}}^{i}\right|^{2}}{\sigma_{k,p,i}^{2}}.\label{eq:def_f}
\end{equation}

We prove that the cross-preamble \ac{cfo} estimator is shown in the
following proposition.
\begin{prop}\label{prop:joint_est_omega}
The cross-preamble \ac{ml} \ac{cfo} estimator is 
\begin{equation}
\hat{\omega}_{k}^{\text{cross}}\triangleq \arg\max_{\omega_{k}}\mathbb{P}\left(\mathbf{r}_{y,k}|\boldsymbol{\omega}\right)=\frac{1}{\tau_{c}}\angle\left\{ \psi_{1}\psi_{0}^{\text{H}}\right\} \label{eq:joint_est_omega}
\end{equation}
where
\begin{equation}
\psi_{0}\triangleq\sum_{p\in\mathcal{P}}\frac{\alpha_{k}^{(p)}r_{y,k}^{p,0}\left[\tau_k\right]^{\text{H}}}{\sigma_{k,p,0}^{2}},\quad\psi_{1}\triangleq\sum_{p\in\mathcal{P}}\frac{\alpha_{k}^{(p)}\mu_{k}r_{y,k}^{p,1}\left[\tau_k\right]^{\text{H}}}{\sigma_{k,p,1}^{2}}.\label{eq:def_phi1}
\end{equation}
\end{prop}
\begin{prop}\label{prop:crlb}
The cross-preamble \ac{ml} \ac{cfo} estimator $\hat{\omega}_{k}^{\text{cross}}\stackrel{a}{\sim}\mathcal{N}(\omega,[\mathbf{I}^{-1}]_{11})$,
where $[\mathbf{I}^{-1}]_{11}$ is \ac{crlb}
\begin{equation}
\left[\mathbf{I}^{-1}\right]_{1,1}=\frac{\Gamma_{0}^{-1}+\Gamma_{1}^{-1}}{2\tau_{\text{c}}^{2}\left|r_{k}\left(\omega_{k}\right)\right|^{2}}\label{eq:crlb_c}
\end{equation}
and $\Gamma_{0}$ and $\Gamma_{1}$ are the aggregated \ac{sinr}
of the two training signals 
\begin{equation}
\Gamma_{0}=\sum_{p\in\mathcal{P}}\frac{\left|\alpha_{k}^{(p)}\right|^{2}}{\sigma_{k,p,0}^{2}},\quad\Gamma_{1}=\sum_{p\in\mathcal{P}}\frac{\left|\mu_{k}\alpha_{k}^{(p)}\right|^{2}}{\sigma_{k,p,1}^{2}}.\label{eq:def_gamma_c}
\end{equation}
\end{prop}

\begin{proof}
(Sketch.) Since the autocorrelation function $r_{k}\left(\omega_{k}\right)$
depends on the \ac{cfo} $\omega_{k}$, we introduce an auxiliary
variable $\gamma_{k}\triangleq r_{k}\left(\omega_{k}\right)$ to absorb
this dependence. Consequently, the unknown parameter vector is reparametrized
as $\boldsymbol{\theta}\triangleq[\omega_{k},\gamma_{k},\gamma_{k}^{\text{H}}]^{\text{T}}$,
where $\omega_{k}$ is real and $(\gamma_{k},\gamma_{k}^{\text{H}})$
is complex. Substituting into the log-likelihood function 
\begin{align*}
f\left(\boldsymbol{\theta}\right) & =\sum_{p\in\mathcal{P}}-\frac{\left|r_{y,k}\left[p,0\right]-\alpha_{k}^{(p)}\gamma_{k}\right|^{2}}{\sigma_{k,p,0}^{2}}\\
 & \quad+\sum_{p\in\mathcal{P}}-\frac{\left|r_{y,k}\left[p,1\right]-\alpha_{k}^{(p)}\gamma_{k}\mu_{k}e^{-j\omega_{k}\tau_{\text{c}}}\right|^{2}}{\sigma_{k,p,1}^{2}}.
\end{align*}

i) \ac{ml} Estimator. Setting the gradients $\partial f/\partial\omega_{k}=0$
and $\partial f/\partial\gamma_{k}^{\text{H}}=0$ yields $\hat{\omega}_{k}^{\text{joint}}=\angle\left\{ \psi_{1}\psi_{0}^{\text{H}}\right\} /\tau_{c}$.

ii) \ac{crlb} Analysis. The \ac{fim} is obtained from the second-order
derivatives of $f\left(\boldsymbol{\theta}\right)$. Taking expectations
over the noise and interference terms, the \ac{fim} is simplified
to
\[
\mathbf{I}=\left[\begin{array}{ccc}
2\tau_{\text{c}}^{2}\Gamma_{1}\left|\gamma_{k}\right|^{2} & j\tau_{\text{c}}\Gamma_{1}\gamma_{k}^{\text{H}} & -j\tau_{\text{c}}\Gamma_{1}\gamma_{k}\\
j\tau_{\text{c}}\Gamma_{1}\gamma_{k}^{\text{H}} & 0 & \Gamma_{0}+\Gamma_{1}\\
-j\tau_{\text{c}}\Gamma_{1}\gamma_{k} & \Gamma_{0}+\Gamma_{1} & 0
\end{array}\right].
\]
By applying the Schur complement, the \ac{crlb} for the \ac{cfo}
is given by 
\begin{equation}
\text{Var}\left(\hat{\omega}_{k}^{\text{joint}}\right)\ge\left[\mathbf{I}^{-1}\right]_{11}=\frac{\Gamma_{0}^{-1}+\Gamma_{1}^{-1}}{2\tau_{\text{c}}^{2}\left|r_{k}\left(\omega_{k}\right)\right|^{2}}\label{eq:crlb}
\end{equation}
\end{proof}

Our analysis of the \ac{crlb} reveals two key insights into \ac{cfo} estimation. First, accurate \ac{cfo} estimation for a weak signal is achievable. This follows from the estimation error being inversely proportional to the aggregate \ac{sinr} across all beams ({\em i.e.}, $\Gamma_{0}^{-1}+\Gamma_{1}^{-1}$). In addition, a coarse-to-fine (iterative) refinement, pre-compensation, and \ac{sic}, followed by re-estimation, can progressively reduce the effective offset because the estimation variance increases with interference and increases with the magnitude of the \ac{cfo}, $|\omega_{k}|$.

\begin{algorithm}[t]
\caption{Joint \ac{cfo} and Channel Estimation Algorithm}
\label{alg:estimation}
\begin{algorithmic}[1]
\item[\#] \textbf{Initialization}: $\hat{\boldsymbol{\omega}}\leftarrow\mathbf{0}$, $\hat{\boldsymbol{\alpha}}\leftarrow\mathbf{0}$, $\hat{\boldsymbol{\mu}}\leftarrow\mathbf{1}$, $\tilde{\mathbf{y}}\leftarrow\mathbf{y}$  
\STATE \label{alg_step:update_channel}Construct $\mathbf{A}$ based on $\hat{\boldsymbol{\omega}}$ and $\hat{\boldsymbol{\mu}}$ and update $\hat{\boldsymbol{\alpha}}$ based on (\ref{eq:est_alpha}); Then, Construct $\mathbf{A}_0$ and $\mathbf{B}$ based on $\hat{\boldsymbol{\omega}}$ and $\hat{\boldsymbol{\alpha}}$ and update $\hat{\boldsymbol{\mu}}$ based on (\ref{eq:est_mu}). 
\STATE For each \ac{bs} $k$, perform \ac{sic} and compensation:  $\tilde{\mathbf{y}}_k\leftarrow(\mathbf{y}-\boldsymbol{A}(\hat{\boldsymbol{\tau}}_{-k},\hat{\boldsymbol{\omega}}_{-k}) \hat{\boldsymbol{\alpha}}_{-k})\varodot \Omega_k(\omega_k)$. Then, estimate the remaining \ac{cfo} $\delta_k$ based on the cross-preamble estimator (\ref{eq:joint_est_omega}), and update $\hat{\omega}_k\leftarrow\hat{\omega}_k+\delta_k$.
\STATE Go to step \ref{alg_step:update_channel}) until $\sum_k\left|\delta_k\right|<\epsilon$.
\end{algorithmic}
\end{algorithm}

\subsection{Joint \ac{cfo} and Channel Estimation Algorithm}
At first, we estimate channels based on (\ref{eq:est_alpha}) and (\ref{eq:est_mu}). Then, for every $k$, we cancel the contributions of the other
\acpl{bs}, pre-compensate the CFO of \ac{bs}~$k$, and estimate its residual CFO
increment $\delta_k$. The to-be-estimated \emph{residual} signal for BS~$k$ is
\begin{equation}
  \tilde{\mathbf{y}}_k
  \;=\;
  \Bigl(\mathbf{y}
  - \boldsymbol{A}\bigl(\hat{\boldsymbol{\tau}}_{-k},\hat{\boldsymbol{\omega}}_{-k}\bigr)
    \hat{\boldsymbol{\alpha}}_{-k}\Bigr)
  \odot \boldsymbol{\Omega}_k(\hat{\omega}_k)
  \label{eq:residual-signal}
\end{equation}
where ``$-k$'' denotes the collection of parameters for all \acpl{bs} except $k$
(i.e., the entries associated with BS $k$ are set to zero). The residual CFO is then updated as
$\hat{\omega}_k \leftarrow \hat{\omega}_k + \delta_k$.
The iteration stops when the residual \ac{cfo} corrections are small.
Intuitively, improving the \ac{cfo} estimate sharpens the matched preamble and
reduces the bias/variance of the channel estimates; the refined channel
estimates, in turn, yield smaller \ac{cfo} increments. Under Proposition~\ref{prop:crlb}, this
yields a monotone decrease of the residual \ac{cfo}, and $\delta_k \to 0$.

\begin{figure}
\begin{centering}
\includegraphics[width=1\columnwidth]{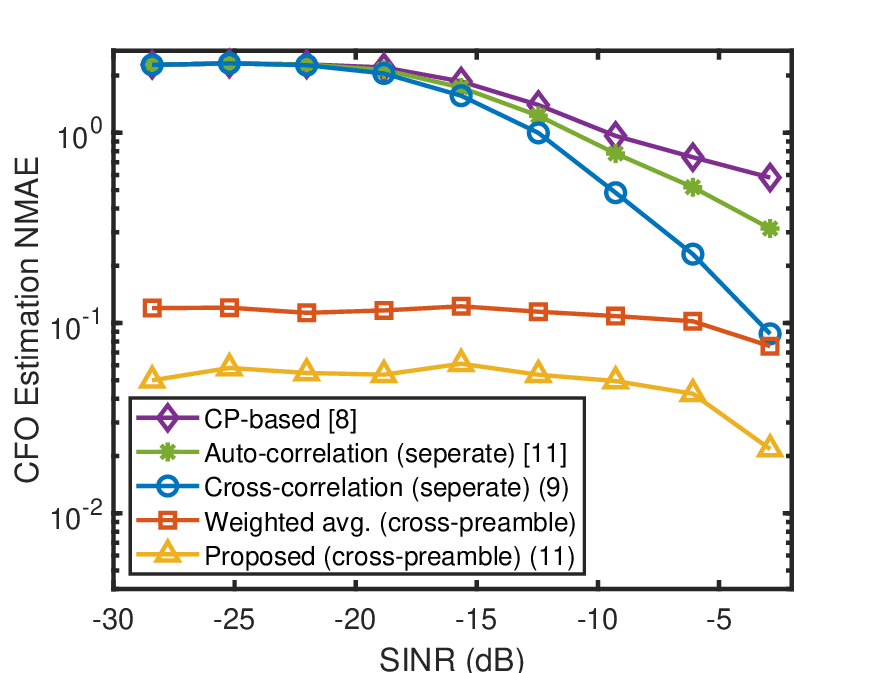}
\par\end{centering}
\caption{\label{fig:cfo_mae_over_sinr}\ac{cfo} $\omega$ estimation NMAE over \ac{sinr},
obtained from 200 Monte-Carlo simulations, with $K=12$ \acpl{bs}
, and $P=12$ preambles.}
\end{figure}

\begin{figure}
\begin{centering}
\includegraphics[width=1\columnwidth]{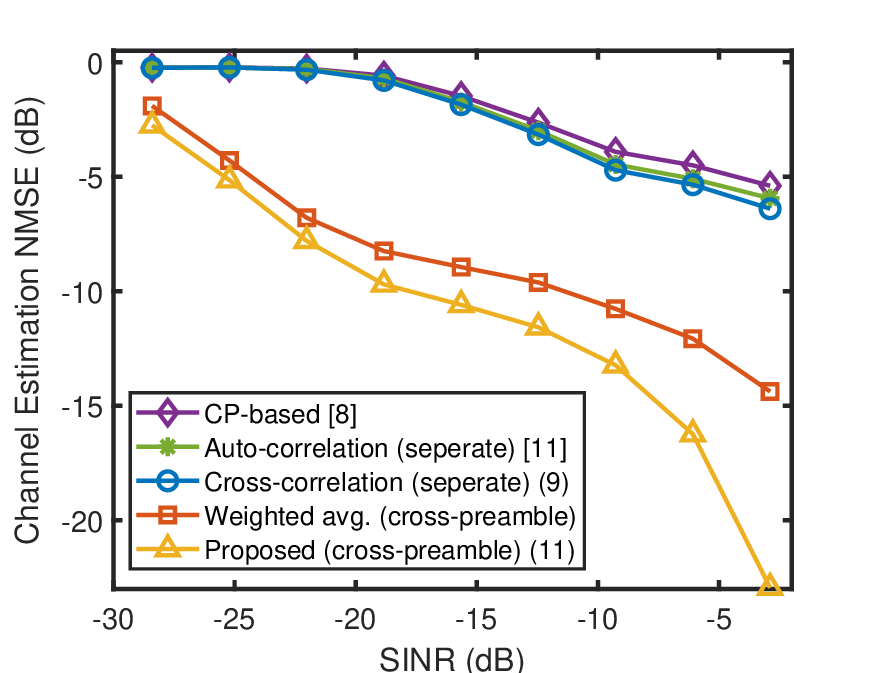}
\par\end{centering}
\caption{\label{fig:alpha_over_sinr}Channel $\alpha$ estimation \ac{nmse} over \ac{sinr},
obtained from 200 Monte-Carlo simulations, with $K=12$ \acpl{bs}
, and $P=12$ preambles.}
\end{figure}

\section{Simulation Results}

We develop a simulation platform that generates 5G \ac{nr} waveforms
\cite{3gpp:ts38211} and use it to assess the robustness of the proposed cross-preamble estimation framework. For comparison, we consider:
i) \ac{cp}-based blind estimator \cite{SinKumMajSat:J25}; ii) Auto-correlation (separate) estimator that splits each training sequence into two segments and estimates \ac{cfo} from their phase difference \cite{TunRivGarMel:J23}; iii) Cross-correlation (separate) estimator as in (\ref{eq:single_est_omega}); iv) Weighted average of separated estimated \acpl{cfo} according to channel power.

Fig.~\ref{fig:cfo_mae_over_sinr} illustrates the \ac{nmae} over \ac{sinr}
\[
\text{NMAE} = \frac{1}{f_{\text{scs}}}\frac{1}{N}\sum_{i=1}^{N}\frac{|\hat{\omega}_i-\omega_i|}{2\pi}
\]
where $f_{\text{scs}}$ is the subcarrier spacing (SCS).

As expected, error decreases with \ac{sinr}. The ranking is consistent with the length of the used reference sequence: cross-preamble \textless{} separate \textless{} \ac{cp}-based. Cross-correlation outperforms auto-correlation because the separation $\tau_{\text{c}}$ between the two training sequences provides a larger phase baseline. The proposed cross-preamble estimator further improves accuracy by pooling information across frames under the shared-\ac{cfo} constraint. Our method achieves a normalized CFO error of approximately 0.05, even at -30 dB SINR. In OFDM systems, keeping the residual CFO well below 0.1 SCS is critical to minimize Inter-Carrier Interference (ICI). This result confirms that our estimator provides sufficient accuracy for reliable symbol demodulation. In contrast, the baseline ("Separate Analysis") degrades significantly as the SINR drops, reaching a normalized error of 1.0 at -15 dB. An error of this magnitude implies a complete synchronization failure (e.g., integer subcarrier shift), rendering subsequent signal recovery impossible. Notably, the cross-preamble estimator at \ac{sinr}$=-30\text{ dB}$ still outperforms the weighted-average baseline operating at \ac{sinr}$=-5\text{ dB}$ confirming robustness in low-\ac{sinr} regimes. Consequently, improving \ac{cfo} accuracy reduces variance in channel estimation; the error (\ac{nmse}) versus \ac{sinr} follows the same trend as in Fig.~\ref{fig:alpha_over_sinr}.

\begin{figure}
\begin{centering}
\includegraphics[width=1\columnwidth]{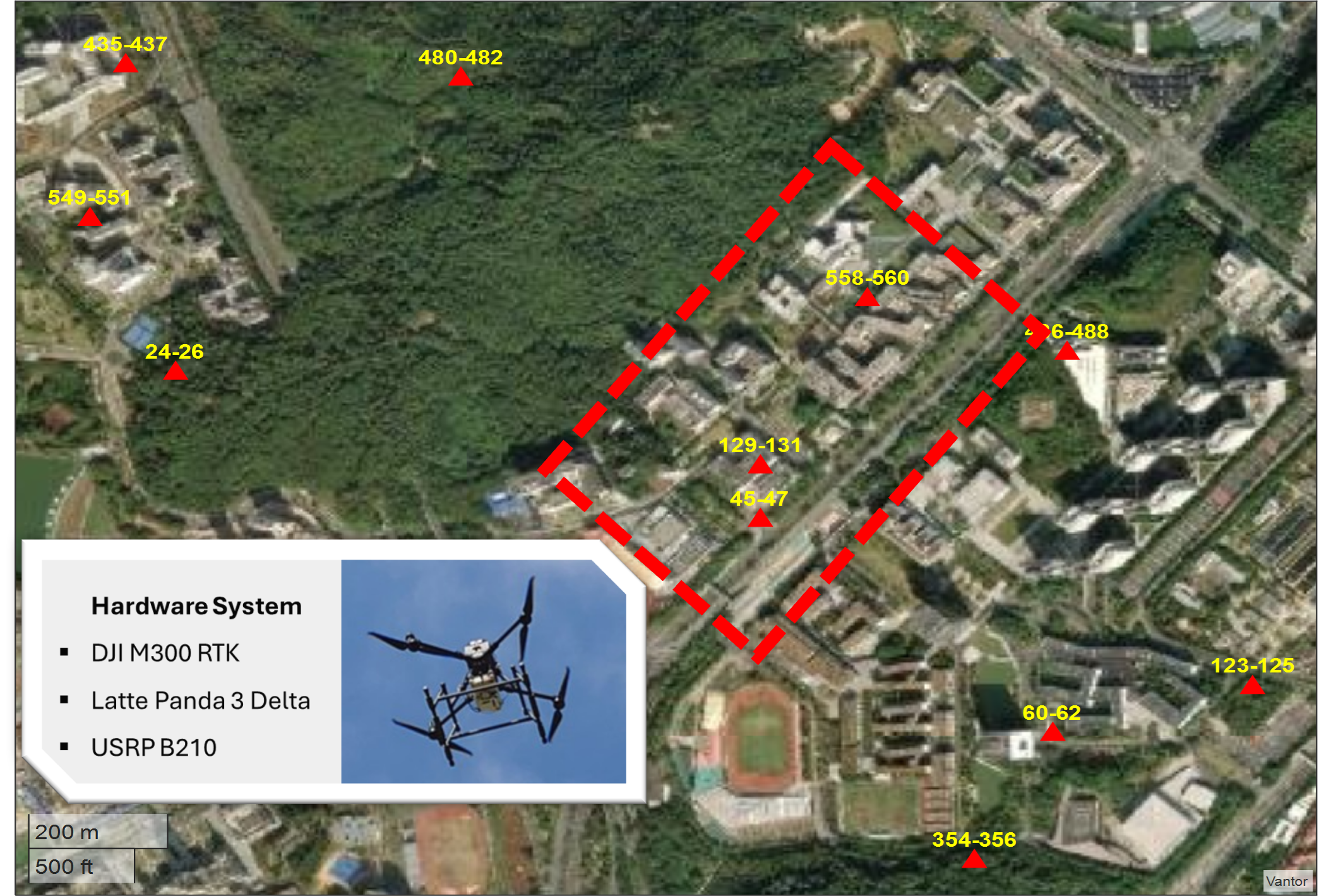}
\par\end{centering}
\caption{\label{fig:test_bed} Aerial sampling campaign over the CUHK-Shenzhen campus. A DJI M300 RTK \ac{uav} equipped with an on-board Latte Panda~3 Delta computer, and a USRP~B210 (calibrated) software-defined radio flies at an altitude of $150\text{ m}$ to collect measurements within the area outlined by the dashed red polygon. Red triangles with yellow labels indicate the locations and Cell ID of a subset of known terrestrial \acpl{bs}.}
\end{figure}

\begin{figure}
\begin{centering}
\includegraphics[width=1\columnwidth]{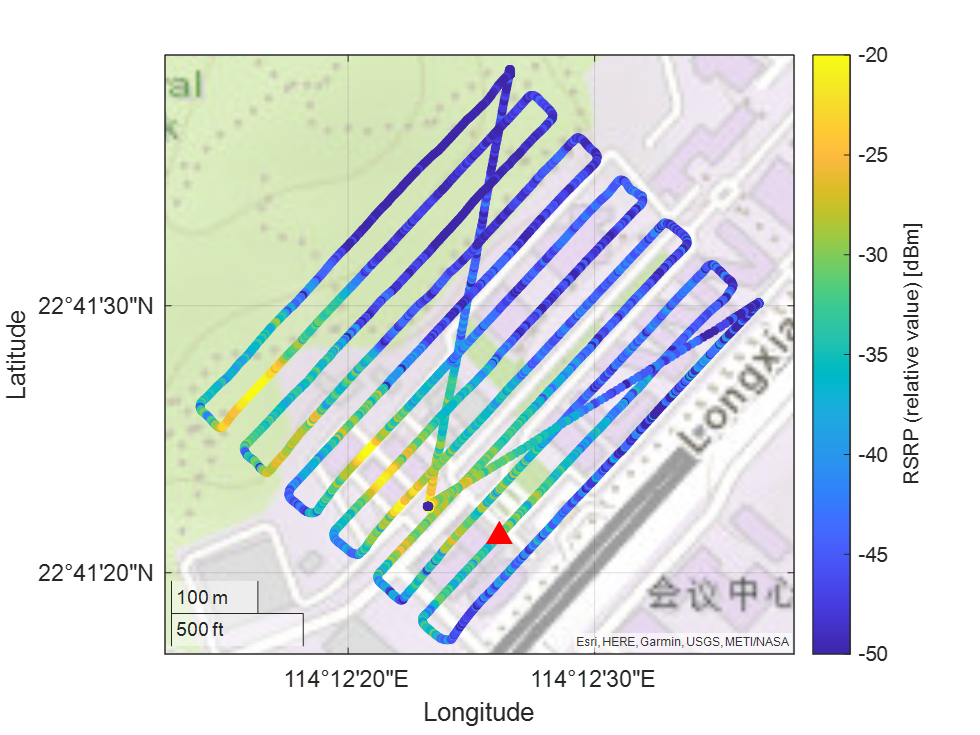}
\par\end{centering}
\caption{\label{fig:radio_map}Field test: Radio maps for beam 0 transmitted by a ground \ac{bs}; the \ac{bs} location is indicated by a red triangle.}
\end{figure}

\begin{figure}
\begin{centering}
\includegraphics[width=1\columnwidth]{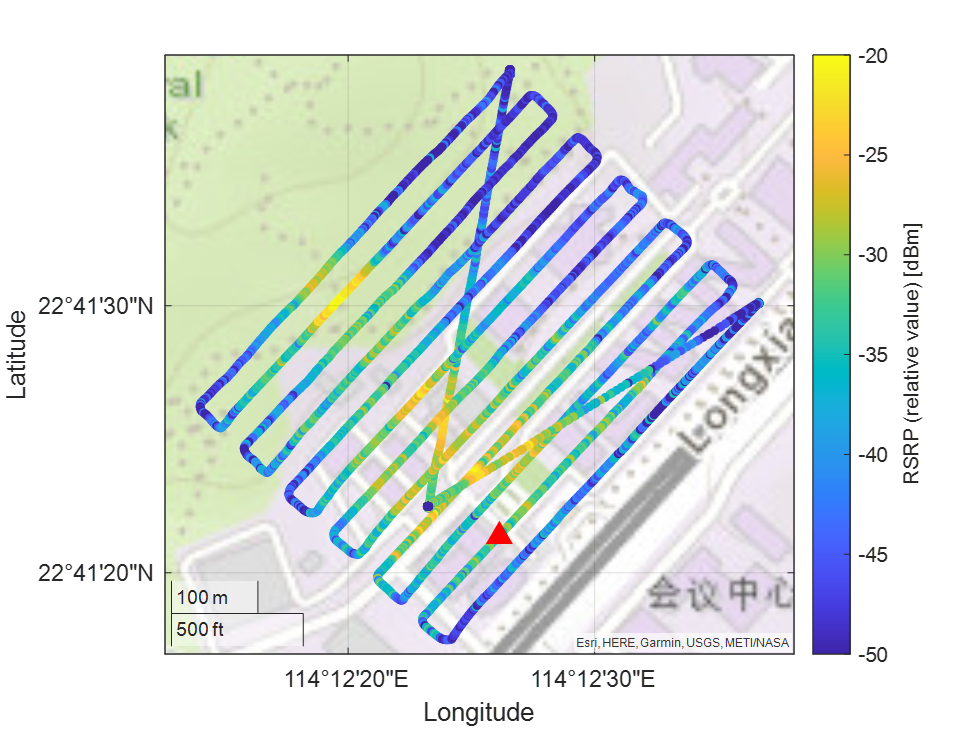}
\par\end{centering}
\caption{\label{fig:radio_map_2}Field test: Radio maps for beam 6 transmitted by a ground \ac{bs}; the \ac{bs} location is indicated by a red triangle.}
\end{figure}

We also verify our algorithm on the testbed, as shown in as illustrated in Fig.~\ref{fig:test_bed}, a DJI M300 RTK \ac{uav} carries a \ac{usrp} to collect 5G \ac{nr} signals, with carrier frequency $f_{c}=2.52495\text{ GHz}$ and bandwidth $20\text{ MHz}$.

Fig.~\ref{fig:radio_map} and \ref{fig:radio_map_2} show the radio-map estimation for a low-altitude setting (2 of 8 beam settings). The two clearly resolved (pointing in different directions) beams corroborate the effectiveness of the proposed framework in practice. It is shown that the received power does not exhibit a simple monotonic distance- or propagation-loss trend but instead shows irregular spatial variations, since the dominant energy in the low-altitude airspace is largely contributed by ground-reflected paths.

\section{Conclusion}

Leveraging \ac{cfo} constancy within a coherence block, we propose an \ac{ml} cross-preamble estimation framework that aggregates preamble correlations. Simulations show estimation gains exceeding $30$\,dB for \ac{cfo} and $5$\,dB for the channel. Field tests demonstrate the ability to identify beams in low-altitude airspace.

\bibliographystyle{IEEEtran}
\bibliography{IEEEabrv,StringDefinitions,BL}

\end{document}

%% file: JCgroup_acronym.tex
\newac{speb}{SPEB}{square position error bound}
\newac[plural=EFIMs,firstplural=Fisher information matrices (EFIMs)]{efim}{EFIM}{Fisher information matrix}
\newac{ne}{NE}{Nash equilibrium}
\newac{mse}{MSE}{mean squared error}
\newac{toa}{TOA}{time-of-arrival}
\newac{snr}{SNR}{signal-to-noise ratio}
\newac{lan}{LAN}{local area network}
\newac{psd}{PSD}{positive semidefinite}
\newac{pd}{PD}{positive definite}
\newac{wrt}{w.r.t.}{with respect to}
\newac{lhs}{L.H.S.}{left hand side}
\newac{wp1}{w.p.1}{with probability 1}
\newac{kkt}{KKT}{Karush-Kuhn-Tucker}
\newac{wlog}{w.l.o.g.}{without loss of generality}
\newac{mle}{MLE}{maximum likelihood estimation}
\newac{gps}{GPS}{global positioning system}
\newac{rssi}{RSSI}{received signal strength indicator}
\newac{mimo}{MIMO}{multiple-input multiple-output}
\newac{csi}{CSI}{channel state information}
\newac{fdd}{FDD}{frequency division duplexing}
\newac{ms}{MS}{mobile station}
\newac{bs}{BS}{base station}
\newac{d2d}{D2D}{device-to-device}
\newac{slnr}{SLNR}{signal-to-interference-leakage-and-noise-ratio}
\newac{ula}{ULA}{uniform linear antenna array}
\newac{pas}{PAS}{power angular spectrum}
\newac{mmse}{MMSE}{minimum mean square error}
\newac{zf}{ZF}{zero-forcing}
\newac{rzf}{RZF}{regularized zero-forcing}
\newac{as}{AS}{angular spread}
\newac{aod}{AOD}{angle of departure}
\newac{iid}{i.i.d.}{independent and identically distributed} 
\newac{sinr}{SINR}{signal-to-interference-and-noise ratio}
\newac{tdd}{TDD}{time-division duplex}
\newac{rvq}{RVQ}{random vector quantization}
\newac{rhs}{R.H.S.}{right hand side}
\newac{mrc}{MRC}{maximum ratio combining}
\newac{cdf}{CDF}{cumulative distribution function}
\newac{a.s.}{a.s.}{almost surely}
\newac{los}{LOS}{line-of-sight}
\newac{jsdm}{JSDM}{joint spatial division and multiplexing}
\newac{map}{MAP}{maximum a posteriori}
\newac{klt}{KLT}{Karhunen-Lo\`eve Transform}
\newac{lbe}{LBE}{link bargaining equilibrium}
\newac{se}{SE}{Stackelberg equilibrium}
\newac{uav}{UAV}{unmanned aerial vehicle}
\newac{nlos}{NLOS}{non-line-of-sight}
\newac{pdf}{PDF}{probability density function}
\newac{em}{EM}{expectation-maximization}
\newac{knn}{KNN}{$k$-nearest neighbor}
\newac{svd}{SVD}{singular value decomposition}
\newac{nmf}{NMF}{non-negative matrix factorization}
\newac{umf}{UMF}{unimodality-constrained matrix factorization}
\newac{rmse}{RMSE}{rooted mean squared error}
\newac{olos}{OLOS}{obstructed line-of-sight}
\newac{mmw}{mmW}{millimeter wave}
\newac{ber}{BER}{bit error rate}
\newac{rss}{RSS}{received signal strength}
\newac{lp}{LP}{linear program}
\newac{ufw}{U-FW}{unimodal Frank-Wolfe}
\newac{utf}{UTF}{unimodality-constrained tensor factorization}
\newac{fw}{FW}{Frank-Wolfe}
\newac{iot}{IoT}{Internet-of-Things}
\newac{mae}{MAE}{mean absolute error}
\newac{crb}{CRB}{Cram\'er-Rao bound}
\newac{aoa}{AoA}{angle of arrival}
\newac{wcl}{WCL}{weighted centroid localization}